\documentclass[superscriptaddress,pre,twocolumn]{revtex4}

\usepackage{graphicx}
\begin{document}
\title{Non-equilibrium dynamics of language games on complex networks}
\author{Luca Dall'Asta}
\affiliation{Laboratoire de Physique Th\'eorique (UMR du CNRS 8627),
B\^atiment 210, Universit\'e de Paris-Sud, 91405 ORSAY Cedex (France)}
\author{Andrea Baronchelli}
\affiliation{Dipartimento di Fisica, Universit\`a ``La Sapienza'' and SMC-INFM,
P.le A. Moro 2, 00185 ROMA, (Italy)}
\author{Alain Barrat} 
\affiliation{Laboratoire de Physique Th\'eorique
(UMR du CNRS 8627), B\^atiment 210, Universit\'e de Paris-Sud, 91405
ORSAY Cedex (France)} 
\author{Vittorio Loreto}
\affiliation{Dipartimento di Fisica, Universit\`a ``La Sapienza'' and SMC-INFM,
P.le A. Moro 2, 00185 ROMA, (Italy)}

\begin{abstract}
  The Naming Game is a model of non-equilibriun dynamics for the
  self-organized emergence of a linguistic convention or a communication
  system in a population of agents with pairwise local interactions. We
  present an extensive study of its dynamics on complex networks, that can be
  considered as the most natural topological embedding for agents involved in
  language games and opinion dynamics. Except for some community structured
  networks on which metastable phases can be observed, agents playing the
  Naming Game always manage to reach a global consensus.  This convergence is obtained
  after a time generically scaling with the population's size $N$ as $t_{conv}
  \sim N^{1.4 \pm 0.1}$, i.e.  much faster than for agents embedded on regular
  lattices.  Moreover, the memory capacity required by the system scales only
  linearly with its size.  Particular attention is given to heterogenous
  networks, in which the dynamical activity pattern of a node depends on its
  degree. High degree nodes have a fundamental role, but require larger memory
  capacity. They govern the dynamics acting as spreaders of
  (linguistic) conventions. The effects of other properties, such as the
  average degree and the clustering, are also discussed.
\end{abstract}

\maketitle

\section{Introduction}\label{sec1}
Understanding the origin and the evolution of language or more
generally of communication systems 
is a fascinating challenge for the interdisciplinary scientific
community~\cite{general}, demanding contributions to researchers in very
different fields, from linguistics to artificial intelligence, from social
sciences to biology, mathematics and physics.  Even though a unitarian view of
language as a complex system is still lacking, a number of different
approaches have been proposed in order to get major insights into some
specific aspects such as, for instance, the self-organized processes leading
to the emergence of a shared lexicon (i.e. a communication system) in a
population of agents.  In the past few years, it has been shown that simple
models of interacting agents could display a 
collective agreement on a shared mapping
between words and objects (or meanings), eventually bootstrapping a shared
system of linguistic conventions, even without global supervision or a priori
common knowledge~\cite{steels1997,kirby2000,barr2004}.  Such models of
``Language Games'', in which the organization of language is tackled at a
purely semiotic level (neglecting semantic relations between symbols and
meanings), have played a pivotal role for the understanding of emergent
communication systems. We will restrain ourselves to the case of the emergence
and evolution of a communication system on short temporal scales compared to
those involved in the evolution of language, so that we neglect any darwinian
principle commonly used in the modelization of language
evolution~\cite{hurford1989,nowak1999}, and focus on simple population
dynamics. In this context, a new field of research called Semiotic Dynamics
has been recently developed, that investigates by means of simple models how
(linguistic) conventions originate, spread and evolve over time in a
population of agents endowed with basic internal states and able to perform
local pairwise interactions~\cite{steels2000}.

The fundamental model of Semiotic Dynamics is the Naming
Game~\cite{steels1995}, that describes a population of agents
trying to agree on the assignation of names to
objects.  The emergence of consensus about the object's name allows
to establish a communication system. Such model was inspired by global
coordination problems in Artificial Intelligence and by peer-to-peer
communication modeling. A practical example of this type of dynamics
is provided by the Talking Heads experiment~\cite{steels1998}, in
which embodied software agents develop their vocabulary observing
objects through digital cameras, assigning them randomly chosen names
and sharing these names in pairwise interactions.  Very
recently~\cite{cattuto2005,huberman}, models of Semiotic Dynamics have
as well found application in the study of a new generation of
web-tools which enable human web-users to self-organize a system of
tags in such a way to ensure a shared classification of information
about different arguments (see, for instance, del.icio.us or
www.flickr.com).

Statistical physics has been involved in the analysis of models of
emergent collective behavior in interacting particles systems for a
long time. It is therefore not a surprise that, recently, various
contributions have come from physicists, in order to shed light on the
dynamics of opinion formation through the study of models of social
interactions~\cite{opinion}. The Naming Game (NG), as a model of
interacting agents reaching a global consensus through emergent
cooperative phenomena, can as well be studied through the statistical
physics toolbox that can give insights into the corresponding complex
dynamics. As a first natural step, previous studies have considered,
as was the case in the Talking Heads experiments, that each agent was
allowed to interact with all the others~\cite{mean-field}. This
mean-field like scenario can indeed be realistic when dealing with a
small number of agents. Moreover, the case of agents embedded into
low-dimensional lattices has as well been investigated~\cite{low-dim},
showing that the global behavior of the Naming Game strongly depends
on the underlying topology. Recently however, the growing field of
complex networks~\cite{bara,mdbook,psvbook} has allowed to obtain a
better knowledge of social networks~\cite{granovetter}, and in
particular to show that the typical topology of the networks on which
agents interact is not regular. The natural step taken in this paper
is thus to consider the Naming Game for agents embedded on more
realistic networks and to study the influence of various complex
topologies on the corresponding dynamical behavior.

The paper is organized as follows. Section~\ref{sec2} and ~\ref{sec3}
are devoted to the definition of the model and to summarize already
known results about the Naming Game dynamics in the case of mean-field
and low dimensional lattices.  In section~\ref{sec4}, we briefly
recall the definition of network models subsequently used in the
theoretical and numerical analysis, whose results are exposed in
section~\ref{sec5}.  Conclusions and directions for future works are
exposed in section~\ref{sec6}.
   
\section{Model definition}\label{sec2} 

\begin{figure} 
\centerline{
\includegraphics*[width=0.45\textwidth]{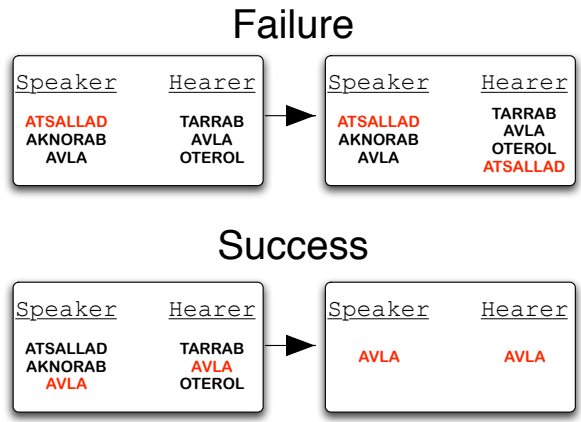}
}
\caption{(Color online) Agents' interaction rules. Each agent 
is described by its inventory, i.e. the repertoire of known words. 
The speaker picks up at random a name in its inventory and transmits it to 
the hearer. If the hearer does not know the selected word the interaction 
is a failure (top), and it adds the new name to its inventory. Otherwise 
(bottom) the interaction is a success and both agents delete all their 
words but the winning one. Note that if the speaker has an empty 
inventory  (as it happens at the beginning of the game), it invents 
a new name and the interaction is, of course, a failure.}
\label{fig:0}
\end{figure}

A minimal model of Naming Game has been put forward by Baronchelli et
al. in Ref.~\cite{mean-field} to reproduce the main features of
Semiotic Dynamics and the fundamental results of adaptive coordination
observed in the Talking Heads experiment~\cite{steels1998}. In this
minimal model, $N$ identical agents observe a single object, for which
they invent names that they try to communicate to one another
through pairwise interactions. Each agent is endowed with an internal
inventory, in which it can store an a priori unlimited number of names
(or opinions). All agents start with empty inventories. At each time
step, a pair of neighboring agents is chosen randomly, one playing as
''speaker'' the other as ''hearer'', and a negotiation process takes
place according to the following rules (see also Figure~\ref{fig:0}).
 The speaker transmits a
name to the hearer. If its inventory is empty, a new word is invented,
otherwise it selects randomly one of the names it knows. If the hearer
has the uttered name in its inventory, the game is a success, and
both agents delete all their words but the winning one. If the hearer
does not know the uttered word, the game is a failure, and the hearer
adds the word to its inventory, i.e. it learns it.
 
%

Note that the time unit is here given by one interaction, in contrast to most
non-equilibrium statistical physics models in which a time unit corresponds to
$N$ interactions. In many cases, results for the dynamical evolution will
therefore be expressed as a function of the rescaled time $t/N$.

Although this model can be seen as belonging to the broad class of
opinion formation models, it is interesting to notice the important
differences with other commonly studied such models~\cite{opinion}.
In particular, each agent can potentially be in an infinite number of
possible discrete states (or words, names), contrarily to the Voter
model in which each agent has only two possible
states. Moreover, an agent can here accumulate in its
memory different possible names for the object, i.e. wait before
reaching a decision, and has an a priori unlimited memory. An
interesting question therefore relates to the actual memory size required
by each agent during the dynamics.  Finally, each dynamical step can
be seen as a negotiation between speaker and hearer, with a certain
degree of stochasticity, while in the Voter model, an agent
deterministically adopts the opinion of one of its neighbors.

Another remark concerns the random extraction of the word to be
uttered from the speaker's inventory. Previously proposed models of
semiotic dynamics used a more complicated representation of the
negotiation interaction assigning weights to the words in the
inventories. In such models (see \cite{lenaerts2005} and references
therein), the word with largest weight is automatically chosen by the
speaker and communicated to the hearer. Communicative success or
failures are translated into updates of the weights: the weight of a
word involved in a successful interaction is increased to the
detriment of the other weights (with no deletion process), while a
failure leads to the decrease of the weight of the word not understood
by the hearer. While such rules are certainly more realistic than the
drastic deletion rule of the minimal Naming Game, the latter has been
shown to retain the essential features of the emergence of a global
collective behavior and corresponds to a much simpler definition.

It is also worth noting that in the minimal Naming Game all agents refer to
the same single object, while in the original experiments the embodied agents
could observe a set of different objects. This is due to the assumption that
homonymy is excluded, i.e. it is impossible that two distinct objects assume
the same name. Thus, in the model, all objects are independent and the general
problem reduces to a set of uncorrelated systems, each one described by the
minimal model.

In the rest of the paper all analysis and numerical simulations will
deal with this simplified model that can be rightly seen as the
prototype of the Naming Game.
 
\section{State of the Art}\label{sec3} 

\begin{figure} 
\centerline{
\includegraphics*[width=0.45\textwidth]{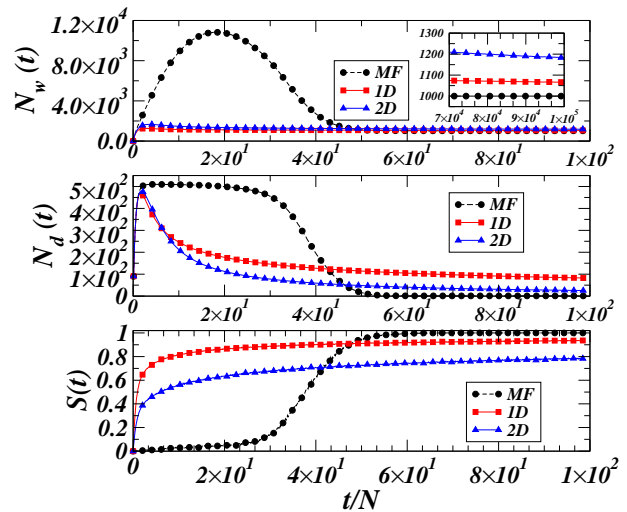}
}
\caption{(Color online) Evolution of the total number of words $N_{w}$
(top), of the number of different words $N_{d}$ (middle), and of the
average success rate $S(t)$ (bottom), for a mean-field system (black circles)
and low dimensional lattices ($1D$, red squares and $2D$, blue
triangles) with $N=1024$ agents, averaged over $10^3$ realizations.
The inset in the top graph shows the very slow convergence for
low-dimensional systems.}
\label{fig:1}
\end{figure}

Most previous studies in Semiotic Dynamics have focused on
populations of agents in which all pairwise interactions are allowed,
i.e. the agents are placed on the vertices of a fully-connected
graph. In statistical mechanics, this topological structure is
commonly referred as ''mean-field'' topology.  In the original work on
the minimal Naming Game model~\cite{mean-field}, Baronchelli et
al. have studied numerically and analytically the behavior of the
mean-field model, providing theoretical arguments in order to explain
the main properties of the population's global behavior. The overall
dynamics have been studied monitoring the temporal evolution of the
total number $N_{w}(t)$ of words in the system at time $t$, i.e. the
total memory used by the agents' inventories, of the number of different
words $N_{d}(t)$, and of the average success rate $S(t)$ (i.e. the
probability, computed averaging on many simulation runs, that the
interaction at time $t$ is successful). At the beginning, many
disjoint pairs of agents interact, with empty initial inventories:
they invent a large number of different words that then start
spreading throughout the system, through unsuccessful interactions.
Indeed, in the early stages of the dynamics, the overlap between
inventories is very low and successful interactions are limited to
those pairs which have been chosen at least twice. Since the number of
possible partners of an agent is of order $N$, an agent rarely
interacts twice with the same partner: the probability of such an
event grows only as $t/N^2$. The consequence, shown in
Fig.~\ref{fig:1} (black circles), is that in this phase the number of
different words $N_{d}$ invented by the agents grows, reaching a
maximum that scales as $\mathcal{O}(N)$. $N_{d}$ then 
saturates (no inventory is empty anymore so that no new
words are invented) and displays a plateau,
while the total number $N_{w}$ of words keep
growing since the various words propagate in the system and
correlations grow between inventories. The peak of $N_{w}$ has been
shown to scale as $\mathcal{O}(N^{1.5})$~\cite{mean-field}, which means
that each agent stores $\mathcal{O}(N^{0.5})$ words, and occurs
after the system has evolved for a time $t_{max} \sim
\mathcal{O}(N^{1.5})$.  The strong correlations built during this time
finally lead the system to consensus in a time $t_{conv}$ of order
$N^{1.5}$. The final state corresponds to the agreement of the agents
on the name to be assigned to a particular object: $N_w=N$, which
means that each agent possesses a unique word in its inventory, and
$N_d=1$, which shows that this word is the same for all agents.  The
S-shaped curve (black circles) of the success rate in
Fig.~\ref{fig:1} summarizes the dynamics: initially, agents hardly
understand each other ($S(t)$ is very low); the inventories start to
present significant overlaps, so that $S(t)$ increases until it
reaches $1$, and the communication system is completely set in.

%

A first study of the effects of topological embedding on the Naming
Game dynamics is reported in Ref.~\cite{low-dim}.  When the
interacting agents sit on the nodes of low-dimensional lattices, the
long-time behavior is still characterized by the convergence to a
homogeneous consensus state, but the evolution of the system changes
considerably.  Since each agent can interact only with a limited
number of neighbors ($2d$ in a $d$-dimensional lattice), at the local
scale the dynamics is very fast: agents can rapidly interact two or
more times with their neighbors, favoring the establishment of a
local consensus with a high success rate (Fig.~\ref{fig:1}, red
squares for $1D$ and blue triangles for $2D$), i.e. of small sets of
neighboring agents sharing a common unique word.  These ''clusters''
of neighboring agents with a common unique word are separated by
individuals having a larger inventory with two or more words, playing
the role of ''interfaces''. For one-dimensional systems, it can be
proved analytically~\cite{low-dim} that the motion of interfaces is a
random walk for which the diffusion coefficient can be computed.
Consequently, the clusters of unique words grow in time with a law
that is typical of coarsening phenomena~\cite{bray}, i.e. the
competition among the clusters is driven by the fluctuations of the
interfaces. The coarsening picture can be extended to higher
dimensions, where it has been checked numerically.  
Such an analysis shows that in
low-dimensional structures local consensus is easy but in the long run
delays the global consensus, which takes much longer to be reached
than in mean-field (see Fig.~\ref{fig:1}): for example, ${\cal
O}(N^3)$ in dimension $1$ vs. ${\cal O}(N^{1.5})$ in mean-field.
However, another important aspect of the problem concerns the memory
used by the agents. In mean-field indeed, each agent needs a memory
capacity scaling as ${\cal O}(N^{1/2})$, i.e. diverging with the
system size. In contrast, the consequence of the embedding in a
finite-dimensional lattice (with finite number of neighbors), and of
the subsequent coarsening like phenomena, with rapid local consensus,
is that each agent uses only a finite capacity: the maximum total
number of words in the system (maximal memory capacity) scales
linearly with the system size $N$ (as for the number of different
words). In summary, {\em low-dimensional lattice systems require more
time to reach the consensus compared to mean-field, but a lower use of
memory}.

Social interactions however take place on networks that are neither
mean-field like nor regular lattices, but share a certain number of
properties such as the small-world property (the average topological
distance between nodes increases very slowly -logarithmically or even
slower- with the number of nodes) and the relative abundance of
``hubs'' (nodes with very large degree compared to the mean of the
degree distribution $P(k)$). More precisely, the degree distributions
are in many cases heterogeneous, with heavy-tails often of a power-law
(or ``scale-free'') nature (for a significant range of values of $k$,
one has $P(k) \sim k^{-\gamma}$)~\cite{bara,mdbook,psvbook}.
Moreover, social networks are often characterized by a large
transitivity, which implies that two neighbors of a given vertex are
also connected to each other with large probability. Transitivity can
be quantitatively measured by means of the clustering coefficient
$c_i$ of vertex $i$ \cite{watts}, defined as the ratio between the
number of edges $m_i$ existing between the $k_i$ neighbors of $i$, and
its maximum possible value, i.e.  $c_i = 2 m_i / (k_i(k_i-1))$. The
average clustering coefficient, defined as $C = \sum_i c_i /N$,
usually takes quite large values in real complex networks.

In order to investigate to what extent these properties can affect the
local and global dynamics of the Naming Game, we have performed
extensive simulations of this model with agents embedded on the nodes
of various paradigmatic computer-generated network models, whose
definitions and main properties are recalled in the next section.

\section{Networks definition}\label{sec4}
 

While many models with various characteristics have been proposed in
the last years in order to account for various detailed properties of
real-world networks, our aim in this paper is to understand the
influence on the dynamics of the Naming Game of the most salient
properties such as heterogeneity in the degree distribution,
clustering, average degree, and we will therefore concentrate on a few
network models that have become indeed paradigms of complex networks.

The prototype of homogeneous networks is the uncorrelated random graph
model proposed by Erd\"os and R\'enyi (ER model)~\cite{erdos}, whose
construction consists in drawing an (undirected) edge with a
fixed probability $p$ between each possible pair out of $N$ given
vertices. The resulting graph shows a binomial degree distribution
with average $\langle k\rangle \simeq Np$, converging to a poissonian
distribution for large $N$. If $p$ is sufficiently small (order
$1/N$), the graph is sparse and presents locally tree-like structures.

In order to account for degree heterogeneity, other constructions have
been proposed for random graphs with arbitrary degree
distributions~\cite{molloy,chung,goh,pastor}. In particular, we will
consider the uncorrelated configuration (UC) model which yields
uncorrelated random graphs through the following construction: $N$
vertices with a fixed degree sequence taken from the desired degree
distribution, with a cut-off $\sqrt{N}$, are connected randomly
avoiding multi-links and self-links.

Since many real networks are not static but evolving, with new
nodes entering and establishing connections to already existing
nodes, many models of growing networks have also been introduced.
We will consider the model introduced by Barab\'asi and
Albert (BA)~\cite{barabasi}, which has become one of the most famous
models for complex heterogeneous networks, and is constructed as
follows: starting from a small set of $m$ interconnected nodes,
new nodes are introduced one by one. Each new node selects $m$
older nodes according to the {\em preferential attachment} rule, i.e.
with probability proportional to their degree, and creates links
with them. The procedure stops when the required network size $N$
is reached. The obtained network has average degree $\langle k \rangle=2m$,
small clustering (of order $1/N$) and a power-law
degree distribution $P(k) \sim k^{-\gamma}$, with $\gamma = 3$.

The BA networks have small clustering, in contrast with social
networks. It turns out that growing networks can as well be
constructed with a large clustering. In Ref.~\cite{dorogovtsev}
indeed, Dorogovtsev et al. have proposed a model (DMS model) in which
each new node connects with the two extremities of a randomly chosen
edge, forming therefore a triangle.  Since the number of edges
arriving to any node is in fact its degree, the probability of
attaching the new node to an old node is proportional to its degree
and the preferential attachment is recovered. The degree distribution
is therefore the same as the one of a BA model with $m=2$, and the
degree-degree correlations are as well equal.  However, the clustering
coefficient is large and approximately equal to
$0.73$~\cite{Barrat2005}. In order to tune the clustering, we can
consider a generalization of this construction, in the spirit of the
Holme-Kim model~\cite{holme2002}: starting from $m$ connected nodes
(with $m$ even), a new node is added at each time step; with
probability $q$ it is connected to $m$ nodes chosen with the
preferential attachment rule (BA step), and with probability $1-q$ it
is connected to the extremities of $m/2$ edges chosen at random
(DMS-like step). The one-node and two-node properties (i.e. degree
distribution and degree-degree correlations) are the same as the
ones of the BA network, while the clustering spectrum, i.e. the
average clustering coefficient of nodes of degree $k$, can be computed
as $C(k)=2(1-q)(k-m)/[k(k-1)]+{\cal O}(1/N)$~\cite{Barrat2005,Barrat2005b}:
changing $m$ and $q$ allow to tune the value of the clustering coefficient.

Since the ER model also displays a low clustering, we consider
moreover a purposedly modified version of this random graph model
(Clustered ER, or CER model) with tunable clustering. Given $N$ nodes,
each pair of nodes is considered with probability $p$; the two nodes
are then linked with probability $1-Q$ while, with probability $Q$, a
third node (which is not already linked with either) is chosen and a
triangle is formed.  The clustering is thus proportional to $Q$ (with
$p\sim {\cal O}(1/N)$ we can neglect the original clustering of the ER
network) while the average degree is approximately given by $\langle k
\rangle \simeq \left[3Q+(1-Q)\right] pN \simeq (2Q+1)pN$
\footnote{Note that, in order to compare an ER and a CER network with
the same $\langle k \rangle$, we therefore tune $p$ for the
construction of the corresponding CER.}.

The next section contains the results of simulations of the minimal
Naming Game with agents embedded on ER and BA networks. Our simulations
have been carried out on networks of sizes ranging from $10^3$ 
to $5.10^4$ nodes, with results averaged over $20$ runs per network
realization and over $20$ network realizations. Since the BA
model has some particular hierarchical structure due to its growing
construction, we have compared the corresponding results with the case
of networks created with the UC model, in which the exponent of the
degree distribution can moreover be varied. It turns out that the
obtained behavior is very similar, so that we will display results
for the BA model. The effect of clustering will be discussed using the
mixed BA-DMS and the CER network models.

\section{Results}\label{sec5}

In this section we expose the main results on the dynamics of the
Naming Game on complex networks. Before entering into the details of
the analysis, it is worth noting that the minimal Naming Game model
itself, as described in section~\ref{sec2}, is not well-defined on
general networks. Indeed, the two neighboring agents chosen to
interact have different roles: one (the speaker) transmits a word and
is thus more "active" than the other (the hearer). One should therefore
specify whether, when choosing a pair, one chooses first a speaker and then
a hearer among the speaker's neighbors, or the reverse order. If the
agents sit on either a fully connected graph or on a regular lattice,
they have an equivalent neighborhood so the order is not important.
On a generic network with degree distribution $P(k)$ however, the
degree of the first chosen node and of its chosen neighbor are
distributed respectively according to $P(k)$ and to $kP(k)/\langle k
\rangle$. The second node will therefore have typically a larger
degree, and the asymmetry between speaker and hearer can couple to the
asymmetry between a randomly chosen node and its randomly chosen
neighbor, leading to different dynamical properties (this is the case
for example in the Voter model, as studied by
Castellano~\cite{castellano}).  This is particularly relevant in
heterogeneous networks for which a neighbor of a randomly chosen node
is a hub with relatively large probability. We therefore can
distinguish more possibilities for the definition of the Naming Game
on generic networks.

\begin{figure} 
\centerline{
\includegraphics*[width=0.45\textwidth]{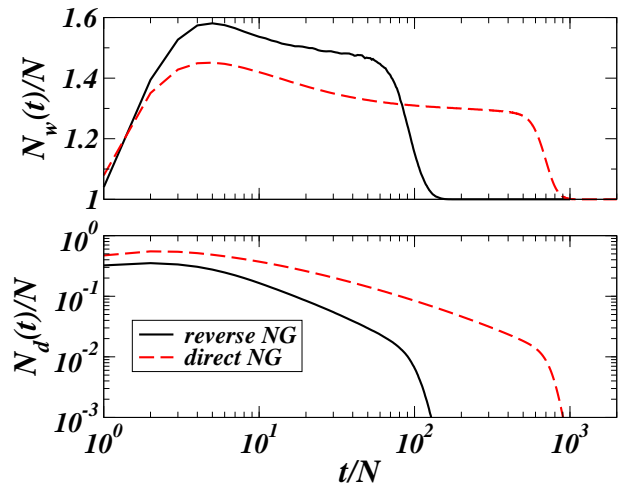}
}
\caption{Total memory $N_{w}$ (top) and number of different words
$N_{d}$ (bottom) vs. rescaled time for two different strategies of
pair selection on a BA network of $N=10^4$ agents, with $\langle k \rangle
=4$.  The reverse NG rule (black full line) converges much faster than
the direct rule (red dashed line). Nonetheless, the two strategies
lead to the same scaling laws with the system size for the convergence
time (not shown).}
\label{fig:2}
\end{figure}

\begin{itemize}
\item (i) A randomly chosen speaker selects (again randomly) a hearer
among its neighbors. This is probably the most natural
generalization of the original rule. We call this strategy {\em
direct Naming Game}. In this case, larger degree nodes will preferentially
act as hearers.

\item (ii) The
opposite strategy, here called {\em reverse Naming Game}, can also be
carried out: we choose the hearer at random and one of its neighbors
as speaker. In this case the hubs are preferentially selected as
speakers.  

\item (iii) A {\em neutral} strategy to pick up pairs of nodes is that
of considering the extremities of an edge taken uniformly at
random. The role of speaker and hearer are then assigned randomly with
equal probability among the two nodes.

\end{itemize}

%
\begin{figure} 
\centerline{
\includegraphics*[width=0.45\textwidth]{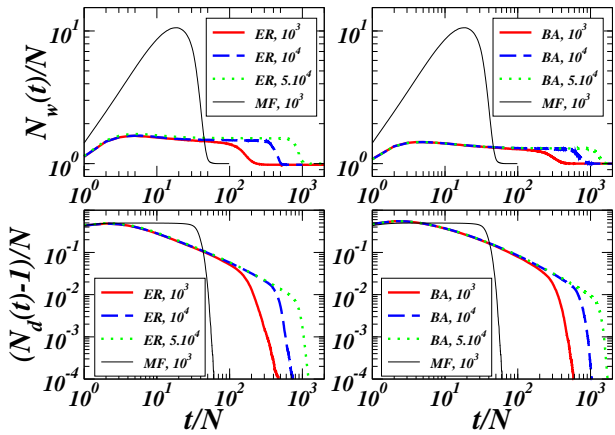}
}
\caption{(Color online) ER random graph (left) and BA scale-free
network (right) with $\langle k \rangle =4$ and sizes $N=10^3, 10^4,
5.10^4$. Top: evolution of the average memory per agent $N_{w}/N$
versus rescaled time $t/N$.  For increasing sizes a plateau
develops in the re-organization phase preceding the convergence.
The height of the peak and of the plateau collapse in this plot,
showing that the total memory used scales with $N$. Bottom: evolution
of the number of different words $N_{d}$ in the system. $(N_{d}-1)/N$
is plotted in order to emphasize the convergence to the consensus with
$N_d=1$. A steady decrease is observed even if the memory $N_w$ displays a
plateau. The mean-field (MF) case is also shown 
(for $N=10^3$) for comparison. }
\label{fig:3}
\end{figure}

Figure~\ref{fig:2} allows to compare the evolution of the direct and the
reverse Naming Game for a BA network of $N=10^4$ agents and $\langle k \rangle
= 4$.  In the case of the reverse rule, a larger memory is used although the
number of different words created is smaller, and a faster convergence is
obtained. This corresponds to the fact that the hubs, playing principally as
speakers, can spread their words to a larger fraction of the agents, and
remain more stable than when playing as hearers, enhancing the possibility of
convergence.  Depending on the network under study, and similarly to the Voter
model case~\cite{castellano}, the scaling laws of the convergence time can
even be modified, as our preliminary study shows. A detailed analysis of this
behavior remains however beyond the scope of our present study and we leave it
for future work (see also~\cite{evolang}). From the point of view of a
realistic interaction among individuals or computer-based agents, the direct
Naming Game in which the speaker chooses a hearer among its neighbors seems
somehow more natural than the other ones. In the remainder of this paper
therefore, we will focus on the direct Naming Game, mentioning where necessary
the corresponding behavior for the other two rules.

%
\begin{figure} 
\centerline{
\includegraphics*[width=0.45\textwidth]{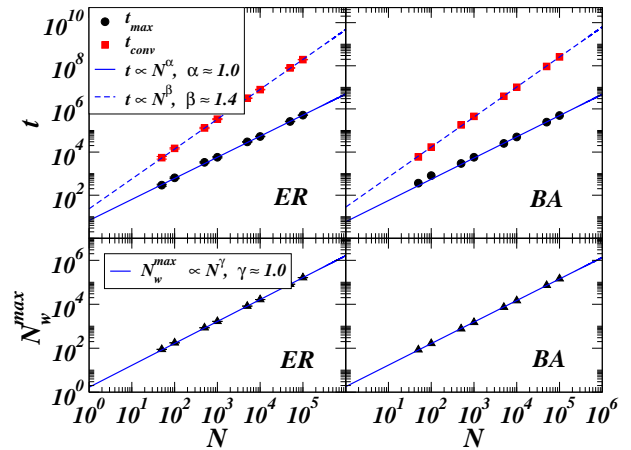}
}
\caption{(Color Online)
Top: scaling behavior with the system size $N$ for the time
of the memory peak ($t_{max}$) and the convergence time ($t_{conv}$)
for ER random graphs (left) and BA scale-free networks (right) with
average degree $\langle k\rangle=4$.  In both cases, the maximal memory
is needed after a time proportional to the system size, while
the time needed for convergence grows as $N^{\beta}$ with $\beta \simeq 1.4$.
Bottom: In both networks the necessary memory capacity (i.e. the maximal
value $N_w^{max}$ reached by $N_w$) scales linearly with the size 
of the network.}
\label{fig:4}
\end{figure}
%
  
\subsection{Global quantities}\label{sec5a} 

We first study the global behavior of the system through the temporal
evolution of three main quantities: the total number $N_{w}(t)$ of
words in the system, the number of different words $N_{d}(t)$, and the
rate of success $S(t)$. All these quantities are averaged over a large
number of runs and networks realizations. In Fig.~\ref{fig:3}, we
report the curves of $N_{w}(t)$ and $N_{d}(t)$ for ER (left) and BA
networks (right) with $N=10^3$, $10^4$ and $5.10^4$ nodes and average
degree $\langle k \rangle = 4$. The corresponding data for the
mean-field case (with $N=10^3$) are displayed as well for
reference. The curves for the average use of memory $N_{w}(t)$ show a
rapid growth at short times, a peak and then a plateau whose length
increases as the size of the system is increased (even when time is
rescaled by the system size, as in Fig.~\ref{fig:3}). The time and
height of the peak, and the height of the plateau, are proportional to
$N$. These scaling properties are systematically studied in
Fig.~\ref{fig:4}, which also shows that the convergence time
$t_{conv}$ scales as $N^{1.4}$ for both ER and BA.  The apparent
plateau of $N_w$ does however not correspond to a steady state, as
revealed by the continuous decrease of the number of different words
$N_d$ in the system: in this re-organization phase, the system keeps
evolving by elimination of words, although the total used memory
almost does not change.

The scaling laws observed for the convergence time is a
general robust feature that is not affected by further topological details,
such as the average degree, the clustering or the particular form of the
degree distribution. We have checked the value of the exponent $1.4\pm 0.1$
for various $\langle k\rangle$, clustering, and exponents $\gamma$
of the degree distribution $P(k)\sim k^{-\gamma}$ 
for scale-free networks constructed with the uncorrelated configuration
model. All these parameters have instead an effect on the other
quantities such as the time and the value of the maximum of memory (see
section~\ref{sec5d}).

\begin{figure} 
\centerline{
\includegraphics*[width=0.45\textwidth]{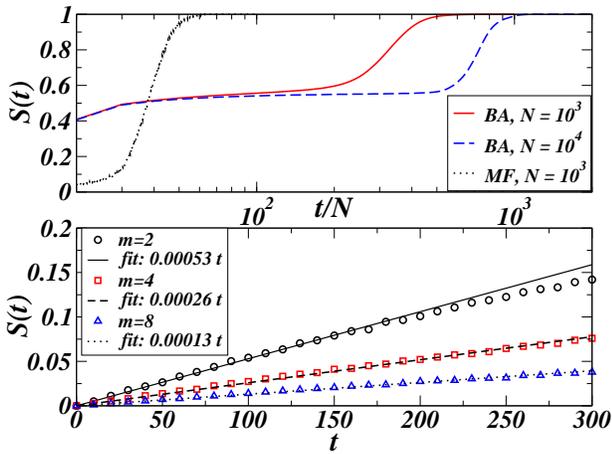}
}
\caption{(Color online) 
Top: Temporal evolution of the success rate $S(t)$ for  
BA scale-free networks  with
$\langle k \rangle =4$ and sizes $N=10^3$ and $10^4$. The dotted black line
refers to the mean-field case ($N=10^3$). The bottom plot
shows the  short time behavior for $N=10^3$ and 
$\langle k \rangle = 4$ ($m=2$, circles),
$\langle k \rangle =8$ ($m=4$, squares), and $\langle k \rangle =16$
($m=8$, triangles). The fitted
slopes are very close to the predicted value $2/\langle k\rangle
N$. 
}
\label{fig:5}
\end{figure}

Figures~\ref{fig:1}, \ref{fig:3} and \ref{fig:4} allow for a direct comparison
between the networks investigated and both the mean-field (MF)
topology and the regular lattices. Thanks to the finite average
connectivity, the memory peak scales only linearly with the system
size $N$, and is reached after a time ${\cal O}(N)$, in contrast with
MF (${\cal O}(N^{1.5})$ for peak height and time) but similarly to the
finite dimensional case. The MF plateau observed in the number of
different words, and corresponding to the building of correlations
between inventories, with an increasing used global memory, and almost
no cancellation of words, is replaced here by a slow continuous
decrease of $N_d$ with an almost constant memory used. With respect to
the slow coarsening process observed in finite dimensional lattices on
the other hand, the small-world properties of the networks, i.e. the
existence of short paths among the nodes, speeds up the convergence
towards the global consensus (see also~\cite{epletter}). Therefore,
complex networks exhibiting small-world properties constitute an
interesting
trade-off between mean-field ''temporal efficiency'' and regular
lattice ''storage optimization''.

Figure~\ref{fig:5} displays the success rate $S(t)$ for 
BA networks with $N=10^3$ (red full line), and $10^4$ (blue
dashed line) agents and $\langle k \rangle =4$. The behaviour for
ER networks is similar. The success rate for
the mean-field ($N=10^3$) is also reported (black dotted lines). 
The success rate increases linearly at very short times
(Bottom plot of Fig.~\ref{fig:5})
then, after a plateau similar to the one observed
for $N_w$, increases on a fast timescale towards $1$. At short times most
inventories are empty, so that the success rate is equal to the
probability that two agents interact twice, i.e. $t/E$, where
$E=N\langle k\rangle/2$ is the number of possible interacting pairs (i.e. 
the number of links in the network), as shown in Fig.~\ref{fig:5}
for BA networks where linear fits to $S(t)$ give slopes in agreement
with the theoretical prediction $2/\langle k\rangle N$. Note that this
argument as well explains that in mean-field the initial success rate
is much lower than for finite $\langle k \rangle$, since there
$E=N(N-1)/2={\cal O}(N^2)$.  When $t \sim \mathcal{O}(N)$, no
inventory is empty anymore, words start spreading through
unsuccessful interactions and $S(t)$ displays a bending.

%
%

\subsection{Clusters statistics}\label{sec5b} 

\begin{figure}
\centerline{
\includegraphics*[width=0.45\textwidth]{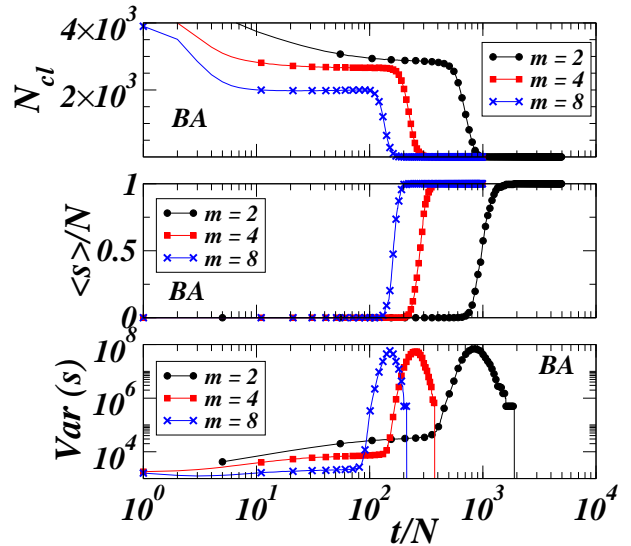}
}
\caption{(Color online)
ER network with $N=10^4$, $\langle k\rangle =4$ (circles), $\langle
k\rangle =8$ (squares), $\langle k \rangle =16$ (crosses). From top to
bottom: Total number of clusters, average normalized cluster size
$\langle s \rangle/N$, fluctuations of the cluster size vs. time.  }
\label{fig:7}
\end{figure}

We now turn our attention to a complementary aspect of the dynamics of
the Naming Game: the behavior of clusters of words.  We call ''cluster''
any set of neighboring agents sharing a common unique word.  In the
case of agents embedded in low-dimensional lattices, it has indeed
been shown~\cite{low-dim} that the dynamics of the Naming Game
proceeds by formation of such clusters, that grow through a coarsening
phenomenon: the average cluster size (resp. the number of clusters)
increases (resp. decreases) algebraically with time. On generic
networks, a different behavior can be expected.  As shown indeed in
Fig.~\ref{fig:7} for the ER model (the behaviour is very similar for the
BA model) the number
of clusters reaches very rapidly a plateau that lasts up to the
convergence time at which it suddenly falls to $1$. Moreover, the
normalized average cluster size remains very close to zero (in fact, of
order $1/N$) during the
plateau, and converges to one with a similar sudden transition. This
transition becomes steeper when the average degree increases (and also
when the size of the system increases), as also emphasized by sharper
peaks in the variance of the cluster size.

%

\begin{figure} 
\centerline{
\includegraphics*[width=0.45\textwidth]{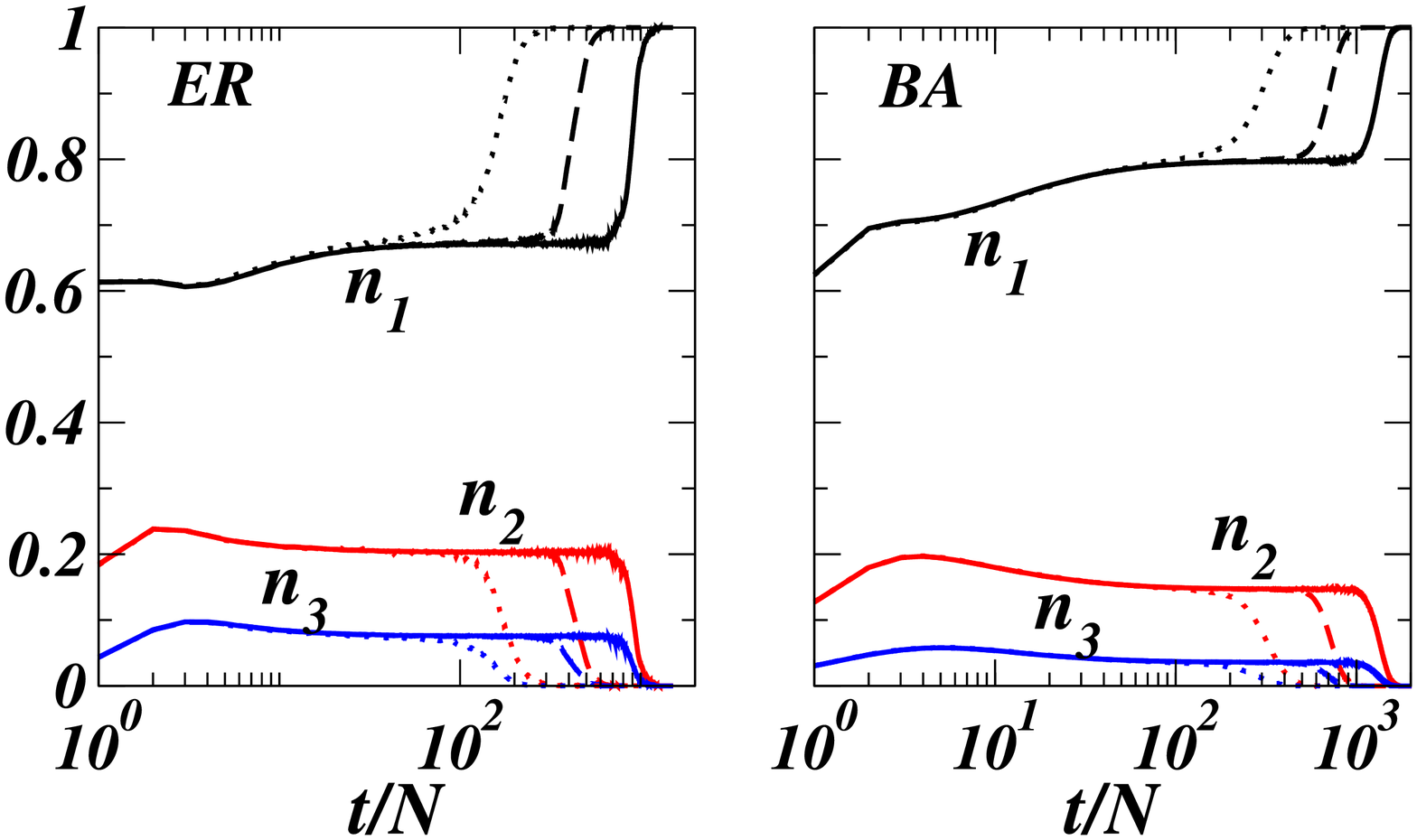}
}
\caption{(Color online)
ER and BA networks with average
degree $\langle k\rangle=4$. Fractions of nodes with one ($n_{1}$ in
black), two ($n_{2}$ in red) and three ($n_{3}$ in blue) words versus
time. These fractions evolve only very slowly during the plateau
displayed by the memory. The curves for three different sizes are
reported: $N=10^3$ (dotted line), $N=10^4$ (dashed line) and
$N=5.10^4$ (full line).  }
\label{fig:9}
\end{figure}

In the same spirit, it is interesting to monitor the number of agents
with a certain number of words: agents with only one word are parts of
clusters while agents using more memory are propagating words from one
part of the system to another.  Fig.~\ref{fig:9} shows the temporal
evolution of the fractions of nodes with $1$, $2$ and $3$
words. As for $N_w$ and the cluster size, these quantities display
plateaus whose length increases with the system size (even in rescaled time
units $t/N$), and converge respectively to $1$ and $0$ abruptly
at $t_{conv}$. Moreover, $n_1$ is much lower than what would be
observed in a coarsening process in which agents with more than one word
are only found at the interfaces.

The emerging picture is very different from the coarsening obtained
on finite dimensional lattices, although the initial formation of small
clusters of agents reaching a local consensus through repeated interactions
is similar. While a majority of nodes soon compose small clusters, the 
fraction of nodes with more words is not negligible and decreases only
at the end of the evolution. Therefore, the dynamics can not be seen
as a coarsening or growth of clusters but as a slow process of
correlations between inventories, in a way much more similar to what
is observed in mean-field~\cite{mean-field}.

\subsection{Effect of the degree heterogeneity}\label{sec5c}

In regular topologies, as well as in mean-field, all agents face an identical
environment. Complex networks are different in that respect, and strong
differences in behavior can be expected for agents sitting on nodes with large
or small degrees. Global properties of dynamical processes are often affected
by the heterogeneous character of the network topology~\cite{mdbook,psvbook}.
The previous subsection however has shown that, similarly to what happens for
the Voter model~\cite{Castellano:2005}, the dynamics of the Naming Game is
similar on heterogeneous and homogeneous networks. Nonetheless, a more
detailed analysis reveals that agents with different degrees present very
different activity patterns, whose characterization is necessary to get
additional insights on the Naming Game dynamics~\cite{inprep}.

%
\begin{figure} 
\centerline{
\includegraphics*[width=0.45\textwidth]{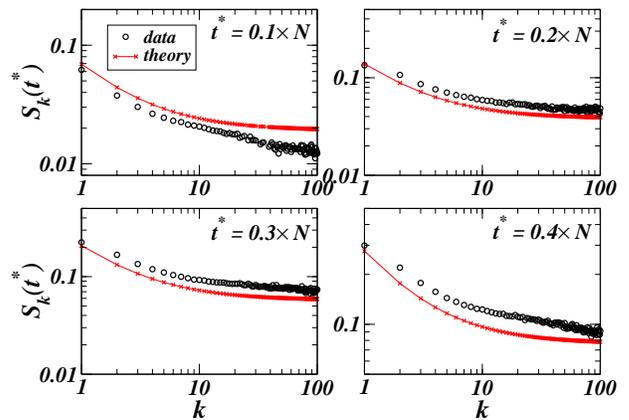}
}
\caption{(Color online)
The four plots correspond to snapshots of the success rate
per degree classes $S_{k} (t^*)$ at times $t^*/N=0.1, 0.2, 0.3, 0.4$ in
a power-law random graph with exponent $\gamma=2.2$ and size $N=10^4$
generated using the uncorrelated configuration model. The numerical
computation of the success rate (black circles) qualitatively agrees
with the numerical evaluation of the expression in Eq.~\ref{rate_k}
(red crosses).  }
\label{fig:10}
\end{figure}

Let us first consider the average success rate $S_{k}(t)$ of nodes
of degree $k$. At the early stages of the dynamics it can be computed
following the arguments of section~\ref{sec5a}. The probability of
choosing twice the edge $i-j$ is
\begin{equation}
\frac{t}{N}\left(\frac{1}{k_{i}}+\frac{1}{k_{j}} \right),
\label{prob_twice}   
\end{equation}
i.e. the probability of choosing first $i$ ($1/N$) then $j$
($1/k_{i}$) or vice versa. Neglecting the correlations between $k_{i}$
and $k_{j}$, one can average over all nodes $i$ of fixed $k_{i} = k$,
obtaining
\begin{equation}
S_{k}(t) \simeq \frac{t}{N}\left( \frac{1}{k} +
\left\langle \frac{1}{k} \right\rangle \right)~.
\label{rate_k} 
\end{equation}
Fig.~\ref{fig:10} shows that, on uncorrelated scale-free networks (UC
model), the data (circles) obtained by numerical simulations are in
qualitative agreement with the direct calculations of the expression
in Eq.~\ref{rate_k} (crosses).  These data together with
Eq.~\ref{rate_k} show that, at the very beginning, the success rate
grows linearly but the effect of the degree heterogeneity is partially
screened by the presence of the constant term $\langle 1/k \rangle$.
The same argument can be used to predict that the success rate should
be essentially degree independent for larger times. $S(t)$ is indeed
always given by two terms, of which only that referring to the node
playing as speaker contains an explicit dependence on $1/k$. The
argument is only approximate since the multiplicative prefactors contain
non-negligible correlations due to the overlapping inventories. More
precisely, these arguments are correct for a neutral Naming Game rule,
but they should hold also for the direct Naming Game in which the constant
term, coming from the activity of nodes as hearers, is much more
relevant for high degree nodes.
 
%
\begin{figure} 
\centerline{
\includegraphics*[width=0.45\textwidth]{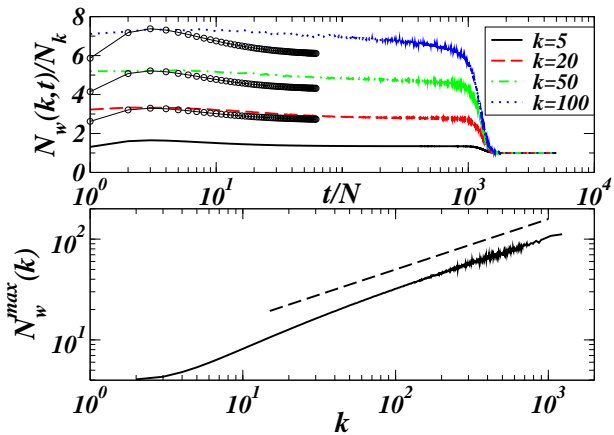}
}
\caption{(Color online)
BA model with $m=2$ (i.e.
$\langle k \rangle=4$), $N=5.10^4$.  Bottom: maximum memory used by a
node as a function of its degree. The dashed line is $\propto
\sqrt{k}$.  Top: average memory used by nodes of degree $k$, for
various values of $k$. The lines show the total memory 
$N_w(k,t)$ used by nodes of degree $k$ at time $t$, normalized by
the number $N_k$ of nodes of degree $k$.
The circles correspond to the bottom curve
($k_0=5$) rescaled by $\sqrt{k/k_0}$ showing the scaling of the peaks.
Note that the values of $N_w(k,t)/N_k$ are averages over many runs
that wash out fluctuations and therefore correspond to smaller values than
the extremal values observed for $N_w^{max}(k)$.  }
\label{fig:12}
\end{figure}

Another interesting point concerns the height of the memory peak. Looking at
classes of nodes of given degree, we get that the height of the memory peak is
larger for nodes of larger degree, as shown in Fig~\ref{fig:12}. This can be 
understood by the fact that hubs act more frequently as hearers and therefore
receive and collect the different words created in the various "areas"
of the network they connect 
together~\footnote{Note that for the reverse Naming Game, the hubs
act more frequently as speakers and therefore accumulate much less
different words. The required maximal memory then slightly decreases
at large $k$.}. In fact, the
maximal memory used by a node of degree $k$ is proportional to $\sqrt{k}$ (see
bottom panel in Fig.~\ref{fig:12}).  For the mean-field case, all agents have
degree $k=N-1$ and the maximal value of the total memory $N_w$ scales indeed
as $N\sqrt{k}=N^{3/2}$. Note however that in the general case, the estimation
of the peak of $N_w$ is not as straightforward. This peak is indeed a
convolution of the peaks of the inventory sizes of single agents, that have
distinct activity patterns and may reach their maximum in memory at different
temporal steps.

\begin{figure}
\centerline{
\includegraphics*[width=0.45\textwidth]{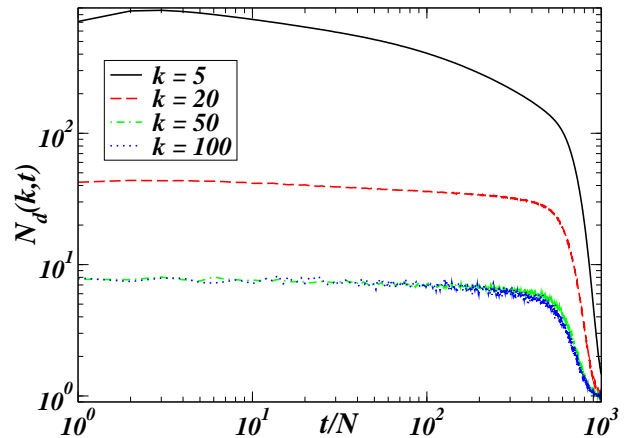}
}
\caption{(Color online)
BA network, $m=2$, $N=10^4$. Total number of different
words in classes of degree $k$, for values of $k=5$, $20$, $50$, $100$
(curves for $k=50$ and $100$ are almost on top of each other).}
\label{fig:13}
\end{figure}

The knowledge of the average maximal memory of a node of degree $k$ is not
sufficient to understand which degree classes play a major role in driving the
dynamics towards the consensus. More insights on this issue can be obtained
observing the behavior of the total number of different words in each degree
class. Figure~\ref{fig:13} shows the evolution of the number $N_{d}(k,t)$ of
different words in the class of nodes with degree $k$, for various values of
$k$ in a BA network with size $N=10^4$ and $\langle k\rangle = 4$. Two
competing effects take part in determining the differences between nodes: high
degree nodes require more memory than low degree nodes (Fig.~\ref{fig:12}),
but their number is much smaller. As a result, low degree classes have in fact
overall a larger number of different words (as shown in Fig.~\ref{fig:13}).
This is due to the fact that during the initial phase, in which words are
invented, low degree nodes are more often chosen as speakers and invent many
different words. The hubs need each a larger memory but they in fact retain a
smaller number of different words.  After the peak in memory, the dynamical
evolution displays a relatively fast decrease of $N_{d}(k,t)$ for small $k$
while a plateau is observed at large $k$: words are progressively eliminated
for low-$k$ nodes while the hubs, which act as intermediaries and are in
contact with many agents, still have typically many words in their
inventories. The role of the hubs, then, is that of diffusing words throughout
the network and their property of connecting nodes with originally different
words helps the system to converge. On the other hand, however, playing mostly
as hearers, the hubs are not able to promote actively successful words, and
their convergence follows that of the neighboring low-degree sites.
In fact, once the low-degree nodes have successfully eliminated most of the
different words created initially, the system globally converges on a faster
timescale. We note that the average memory $N_w(k,t)/N_k$ converges slightly
faster than $N_d(k,t)$ (and that $N_d(k,t)$ converges faster for larger $k$),
showing that the very final phase consists in the late adoption of the
consensus by the lowest degree nodes, in a sort of final cascade from the
large to the small degrees.

\subsection{Effect of the average degree and clustering}\label{sec5d}
 
Social networks are generally sparse graphs, but their structure is
often characterized by high local cohesiveness, that is the result of
a very natural transitive property of many social
interactions~\cite{granovetter}.  The simplest way to take into
account these features on the dynamics of Naming Game is that of
studying the effects of changing the average degree and the clustering
coefficient of the network.

\begin{figure} 
\centerline{
\includegraphics*[width=0.45\textwidth]{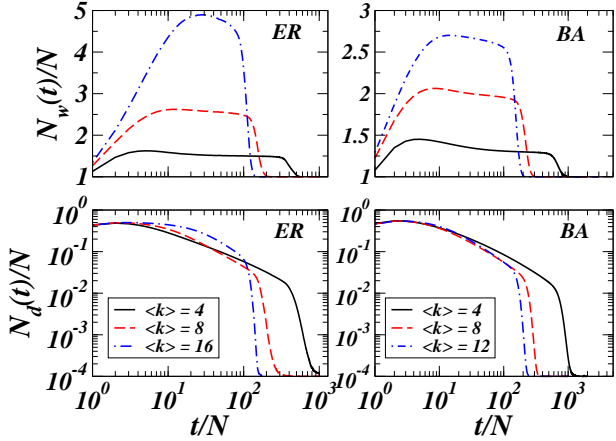}
}
\caption{(Color online)
ER networks (left) and BA networks (right) with $N=10^4$
agents and average degree $\langle k \rangle=4$, $8$, $16$. The increase of
average degree leads to a larger memory used ($N_w$, top)
but a faster convergence. 
The maximum in the number of different words is not affected by
the change in the average degree (bottom). }
\label{fig:14}
\end{figure}

%

The effects of increasing the average degree on the behavior of the
main global quantities are reported in Fig.\ref{fig:14}. In both ER
(left) and BA (right) models, increasing the average degree provokes
an increase in the memory used, while the global convergence time is
decreased. 
%
%
Note also that, while the behavior of the convergence time with $N$
(i.e. a power-law $N^\beta$ with $\beta \approx 1.4$) is very robust,
the linear scaling for the memory peak properties ($N_w^{max} \propto
N^\alpha$ and $t_{max}\propto N^\alpha$ with $\alpha=1$), are slightly altered
by an increase in the average degree (not shown).
Increasing $\langle k\rangle$ at finite $N$ brings indeed the system closer to
the mean-field behavior where the scaling of these quantities is
non-linear ($\alpha_{MF}=1.5$); at large enough sizes however,
the linear scaling is recovered.

Moreover, for larger average degree,
the number of nodes having only one word decreases (not shown); i.e. the system
needs a more complicated re-organization phase that involves a larger
number of agents with many words, but induces a faster convergence.
In fact, the larger possibilities of interaction given by the
larger number of connections allows for a better sharing of common words
and for a more efficient correlation of inventories,
thus favoring a faster convergence. 


\begin{figure} 
\centerline{
\includegraphics*[width=0.45\textwidth]{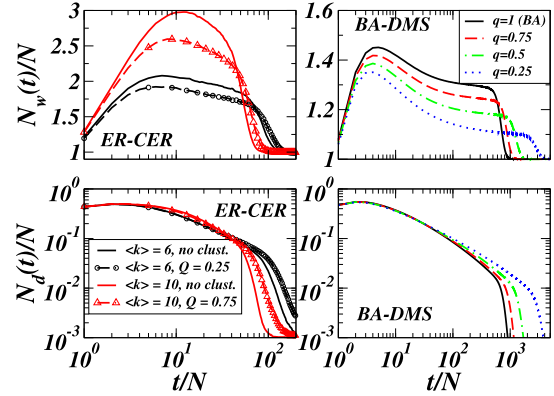}
}
\caption{(Color online)
Effect of clustering on the behavior of the total number of
words $N_{w}(t)$ and of the number of different words $N_{d}(t)$ on
random graphs (left) and scale-free networks (right) with $N=10^4$.
In order to inject triangles into the ER random graphs we have used the
construction described in section~\ref{sec4}, obtaining clustered
random graphs (CER model, with clustering coefficient proportional to
$Q$) that have been compared to standard ER graphs with equal
average degree ($\langle k\rangle = 6$ and $10$). Scale-free networks
have been generated with the mixed BA-DMS model described in
section~\ref{sec4}, in which the clustering coefficient is
proportional to $1-q$. In both networks higher clustering leads to
smaller memory capacity required but a larger convergence time.}
\label{fig:16}
\end{figure}
%
\begin{figure} 
\centerline{
\includegraphics*[width=0.45\textwidth]{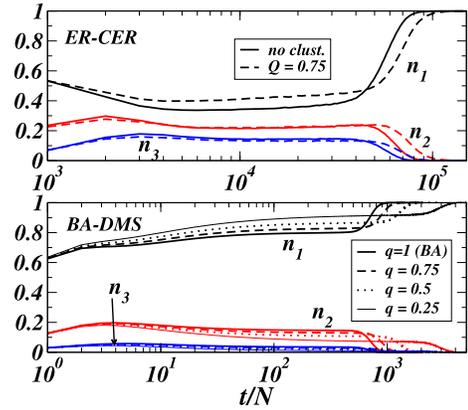}
}
\caption{(Color online)
Effect of enhanced clustering on the fraction of agents with $1$
($n_{1}$ in black), $2$ ($n_{2}$ in red) and $3$ ($n_{3}$ in blue)
words.  Top: We compare a clustered
random graph (CER model, with clustering coefficient proportional to
$Q$) to standard ER graphs, both with average degree $\langle
k\rangle = 10$.  Bottom: Scale-free networks have been generated with the
mixed BA-DMS model described in section~\ref{sec4}, in which the
clustering coefficient is proportional to $1-q$.  In both cases there
is a tendency to increase the fraction of agents with one word
and decrease the others fractions. } 
\label{fig:17}
\end{figure}
%

Note that the clustering is slightly changing when changing the
average degree, but its variation is small enough for the two effects
to be studied separately. Here we use some other mechanisms to enhance
clustering, summarized in the following two models that
have been defined in section~\ref{sec4}: clustered Erd\"os-R\'enyi
(CER) random graphs, and mixed BA-DMS model. 

Figure~\ref{fig:16} shows the effect of increasing the clustering at
fixed average degree and degree distributions: the number of different
words is not changed, but the average memory used is smaller and the
convergence takes more time. Moreover, the memory peak at fixed $k$ is
smaller for larger clustering (not shown): it is more probable for a
node to speak to $2$ neighbors that share common words because they
are themselves connected and have already interacted, so that it is
less probable to learn new words. Favored by the larger number of
triangles, cliques of neighboring nodes learn from the start the same
word, causing a slight increase in the fraction of nodes with only one
word as reported in Fig.~\ref{fig:17} for both homogeneous and
heterogeneous networks. At fixed average degree i.e. global number of
links, less connections are moreover available to transmit words from
one part of the network to the other since many links are used in
``local'' triangles. The enhanced local coherence therefore is in the
long run an obstacle to the global convergence. We note that this
effect is similar to the observation by Holme et
al.~\cite{holme2002b} that, at fixed $\langle k \rangle$, more clustered
networks are more vulnerable to attacks since many links are ``wasted''
in redundant local connections.

\begin{figure} 
\centerline{
\includegraphics*[width=0.45\textwidth]{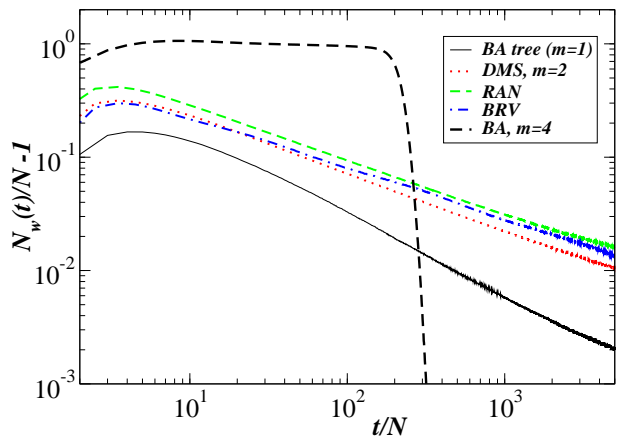}
}
\caption{Evolution of the number of words for the Naming Game 
on hierarchical networks, namely the BA tree, the DMS model,
the hierarchical BRV~\cite{Barabasi:2001}, and the Random
Apollonian network. The case of the BA model with $m=4$ is shown
with a dashed line for reference.}
\label{fig:18}
\end{figure}
%
\subsection{Effect of hierarchical structures}\label{sec5e}

In the previous sections we have argued that networks with small-world
property have fast (mean-field like) convergence after a re-organization phase
whose duration depends on other properties of the system. The
small-world property holds when the diameter of the network 
grows slowly, i.e. logarithmically or slower, with its size $N$.
This ensures that every part of the network is rapidly reachable from
any other part, in contrast to what happens with regular lattices.
Such property therefore generically enhances the possibility
of creating correlations between the inventories of the agents
and of finally converging to a consensus.
In this subsection, we show that this line of reasoning bears, 
surprisingly at first sight, some exceptions. 

The first (and easiest to apprehend) exception is given by the scale-free
trees, obtained by the preferential attachment procedure with $m=1$. In this
case, as shown in Fig.~\ref{fig:18}, the convergence is reached very slowly,
with $N_w(t)/N -1$ decreasing as a power law of the time. This is in contrast
with the generic behavior, i.e.  a plateau followed by an exponential
convergence, as shown also for reference in Fig.~\ref{fig:18}, but similar to
the finite-dimensional lattices (the average cluster size as well grows as a
power-law, in contrast with the data of Fig.~\ref{fig:7}). This is
reminiscent of what happens for the Voter model \cite{Castellano:2005}, in
which a power-law instead of exponential decrease of the fraction of active
bonds is observed, and can be understood through the tree structure of the
network.  Indeed, from the viewpoint of the dynamics, a tree is formed by two
ingredients: linear structures on which the interfaces between clusters
diffuse as in one-dimensional systems and branching points at which interfaces
may be pinned and their motion slowed. In fact, we have checked that similar
(slow) power-law behaviors are also obtained for the Naming Game on Cayley
tree (i.e. in which every node has the same degree) or for scale-free tree
with different degree distributions (obtained through the generalized linear
preferential attachment model).

The slowness of this dynamical behavior is however rooted in a slightly more
subtle consideration. As Fig.~\ref{fig:18} indeed shows, the Naming Game
displays power-law convergence in other heterogeneous networks that are not at
all tree-like, such as the DMS model with $m=2$~\cite{dorogovtsev}, in which
at each step a triangle is created, the deterministic scale-free networks of
Barab\'asi, Ravasz and Vicsek (BRV)~\cite{Barabasi:2001} or the Apollonian and
Random Apollonian Networks (RAN)~\cite{Andrade:2005,Zhou:2005}. Let us briefly
recall how these networks are constructed:
\begin{itemize}
\item For the DMS model with $m=2$, one adds at each step a new node which is
  connected to the extremities of a randomly chosen edge. 
\item For the deterministic scale-free BRV networks,
  one starts (step $1$) with two nodes connected to a root. At each step $n$,
  two units (of $3^{n-1}$nodes) identical to the network formed at the
  previous step are added, and each of the bottom $2^n$ nodes are connected
  to the root.  
\item The Random Apollonian networks are embedded in a two-dimensional plane.
  One starts with a triangle; a node is added in the middle of the triangle
  and connected to the three previous nodes; at each step, a new node is added
  in one of the existing triangles (chosen at random) and connected to its
  three corners, replacing the chosen triangle by three new smaller triangles.
\end{itemize}
All these networks share a very important and hard to quantify property: they
are hierarchically built. This is particularly clear for the BRV case, since
at each step the new network is formed by three identical sub-networks. In the
RAN as well, hierarchically nested units can be identified with the triangles,
each of which contains other smaller triangles. Finally, in the DMS case, one
can identify a unit at a certain scale as an edge and the set of nodes that
have been attached to the extremities of this edge or of the edges
subsequently created in this unit.
Because of these particular network organizations, each node belongs in fact
to a given sub-hierarchical unit and, to go from one node to another node in
another sub-unit, a hierarchical path has to be followed. The trees represent
a particular class of such structures, in which there exist only one path
between two given nodes.  In this sense, such networks, although being
small-world, present a structure which renders communication between different
parts of the network more difficult.  Each sub-unit can therefore converge
towards a local consensus which renders the global consensus more cumbersome
to achieve.

Such results show that the small-world property in fact does not by itself
guarantee an efficient convergence of dynamical processes such as the Naming
Game, and that strongly hierarchical structures in fact slow down
and obstruct such convergence.

\section{Discussion and Outlook}\label{sec6}

In this article, we have studied the dynamical properties of the
minimal Naming Game model~\cite{mean-field} in populations of
agents interacting on complex topologies, focusing on homogeneous
and heterogeneous networks (represented respectively by the 
Erd\"os-R\'enyi random graph model and by the Barab\'asi-Albert
scale-free model). Social networks indeed are typically neither fully
connected graphs nor regular lattices. We have considered the effects
of various networks characteristics such as the heterogeneity, the average
degree and the presence of clustering.

The main characteristic of the studied networks is the small-world
property (the average hopping distance between two nodes scales only
logarithmically with the size of the network). After an initial phase
during which words are created, the small-world property ensures their
propagation out of the local scale, boosting up the spreading process
contrarily to what happens in low dimensional lattices where words'
spreading is purely diffusive (see sections~\ref{sec5a}-\ref{sec5b}).
As already suggested in Ref.~\cite{epletter}, we argue that the
small-world property allows to inhomogeneous and sparse networks to
recover the high temporal efficiency observed in the mean-field
system. For both the ER and BA network models we get a scaling law for
the convergence time $t_{conv}$ with the size $N$ of the system of the
type $t_{conv} \sim N^{\beta_{SW}}$, with exponent approximately
$\beta_{SW} \simeq 1.4$.  The discrepancy with the mean-field exponent
($\beta_{MF} \simeq 1.5$) may be due to logarithmic corrections that
are unlikely to be captured using numerical scaling
techniques. Moreover, small-world networks have higher memory
efficiency than the mean-field model, since the peak in the total
number of words scales only linearly with the size $N$. This is due to
the fact that these networks are {\em sparse}, i.e. their average
degree $\langle k\rangle$ is small compared to $N$.

The detailed analysis of the Naming Game dynamics shows distinct activity
patterns on homogeneous and heterogeneous networks.  In homogeneous networks
all nodes have a similar neighborhood and therefore similar dynamical
evolution, while in heterogeneous networks classes of nodes with different
degree play different roles in the evolution of the Game. The role of the hubs
is better understood thanks to the degree based analysis of the number of
words and different words. High degree nodes, indeed, are more likely chosen
as hearers and, consequently, they have larger inventory sizes.
 At the beginning, because of the pair choosing
strategy (direct Naming Game), low degree nodes are much more involved in the
process of word generation than the hubs. Local consensus is easily reached
and a large amount of locally stable different words get in touch with higher
degree nodes. The latter start to accumulate a large number of words in their
inventories, playing as spreaders of names towards less connected
agents and finally driving the convergence.
From this viewpoint, the convergence dynamical pattern of the Naming Game on
heterogeneous complex networks presents some similarities with more studied
epidemic spreading phenomena~\cite{barthelemy}. A more detailed comparison of
the activity pattern for the direct and reverse Naming Game is left for future
work~\cite{inprep}.

The relation between topological properties and dynamical evolution of the
system are further characterized by a detailed study of the effects of varying
the average degree and clustering coefficient.  These effects are equivalent
on homogeneous and heterogeneous networks. While any increase of the average
degree provokes a larger memory peak and a faster convergence, the growth of
clustering coefficient leads to the decrease of the necessary memory but the
fast obtention of local consensus delays in the long run the global
convergence.  The latter effect is particularly relevant for real social
networks in which local cohesiveness is an important feature that cannot be
neglected. Another important ingredient of real networks that we have not
addressed here is the presence of degree correlations in the network topology.
It would indeed be interesting to know in what measure positive or negative
degree correlations affect the negotiating processes of the agents.

\begin{figure} 
\centerline{ \includegraphics*[width=0.45\textwidth]{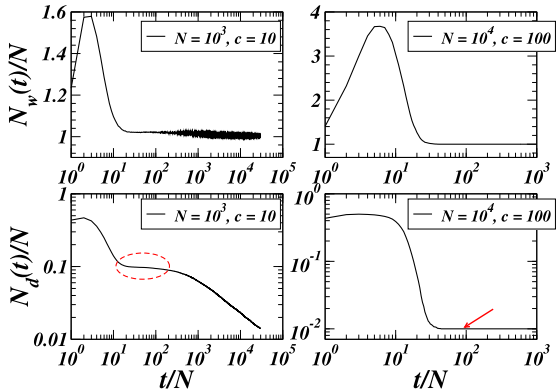}
}
\caption{Metastable states in networks with strong community structure.
Each community is composed of $c$ nodes so that there are $N/c=100$
communities. The dashed ellipse and the arrow identify the metastable
states with $N_d=N/c$ different words.}
\label{fig:19}
\end{figure}

In summary, as other models of opinion formation~\cite{opinion}, the Naming
Game shows a non-equilibrium dynamical evolution from a disordered state to a
state of global agreement. However, with respect to most opinion models, in
which the agents may accept or refuse to conform to the opinion of someone
else, the Naming Game gives more importance to the bilateral negotiation
process between pairs of agents that is a cornerstone in the evolutionist
theory of Language~\cite{general}. For this reason the Naming Game should be
regarded as an independent attempt to model the ultimate emergence of a
globally accepted linguistic convention or, in other terms, the establishment
of a self-organized communication system.  In contrast with other
non-equilibrium models, as those based on zero-temperature Glauber dynamics or
the voter model~\cite{block,Castellano:2005}, we do not find any signature of
the occurrence of metastable blocked states in any relevant topology with
quenched disorder.  While the total number of words displays a plateau whose
length increases with the system size during the re-organization phase,
indeed, the number of different words is continuously decreasing, revealing
that the convergence is not a matter of fluctuations due to finite-size
effects, but the result of an evolving self-organizing process.  Such behavior
makes the Naming Game a robust model of self-coordinated communication in any
structured population of agents.  A noticeable exception concerns the case of
agents sitting on networks with strong community structures, i.e.  networks
composed of a certain number of internally highly connected groups
interconnected by few links working as bridges.  Figure~\ref{fig:19} reports
the behavior of the Naming Game on such a network, composed of fully-connected
cliques, each of $c$ nodes, the various cliques being connected to each other
with only one link.  From simulations it turns out that, not only the total
number of words, but even the number of different words display a plateau
whose duration increases with the size of the system. The number of different
words in the plateau equals the number of communities, while the corresponding
total number of words per node is about one, proving the existence of a real
metastable state in which the system reaches a long-lasting multi-vocabulary
configuration. Indeed, each community reaches internal consensus but the weak
connections between communities are not sufficient for words to propagate from
one community to the other.

%

In conclusion, populations of agents with fixed complex topology do evolve
towards a homogeneous state of consensus and efficient communication, except
for somehow artificial network structures, the detailed topological properties
affecting only the convergence pattern and time scale. Future work will
address the important issue of a possible interplay between topology and
dynamics in populations in which the agents are free of rearranging their
connectivity patterns in relation to local (or global) information on the
dynamical evolution of the system. It would for example be interesting to
verify if such interplay may allow for a natural emergence of community
structures and multi-language cohabitation.
 
\section{Acknowledgments} 
The authors thank E. Caglioti, C. Cattuto and L. Steels for many
useful discussions.  A. Baronchelli and V. L. are partially
supported by the EU under contract IST-1940 (ECAgents).
A. Barrat and L.D. are partially supported by the EU under contract
001907 ``Dynamically Evolving, Large Scale Information Systems''
(DELIS).

\end{document}